\documentclass[mmnp]{edpsmath}
\usepackage{graphicx}
\usepackage{amsmath}
\usepackage{amssymb}

\begin{document}
\title{On stability of linear dynamic systems with hysteresis feedback}
%
\author{Michael Ruderman}
\address{Faculty of Engineering and Science,
University of Agder, p.b. 422, Kristiansand, 4604-Norway. \\email:
michael.ruderman@uia.no}
\date{October 2018}

\begin{abstract}
The stability of linear dynamic systems with hysteresis in
feedback is considered. While the absolute stability for
memoryless nonlinearities (known as Lure's problem) can be proved
by the well-known circle criterion, the multivalued
rate-independent hysteresis poses significant challenges for
feedback systems, especially for proof of convergence to an
equilibrium state correspondingly set. The dissipative behavior of
clockwise input-output hysteresis is considered with two boundary
cases of energy losses at reversal cycles. For upper boundary
cases of maximal (parallelogram shape) hysteresis loop, an
equivalent transformation of the closed-loop system is provided.
This allows for the application of the circle criterion of
absolute stability. Invariant sets as a consequence of hysteresis
are discussed. Several numerical examples are demonstrated,
including a feedback-controlled double-mass harmonic oscillator
with hysteresis and one stable and one unstable poles
configuration.
\end{abstract}
\begin{resume} ... \end{resume}

\subjclass{XXX}

\maketitle


\section{Introduction}
\label{sec:1}

The stability of dynamic systems containing hysteresis has been
investigated for some time, but still remains a challenging and
appealing topic for systems and control theory. Much the same can
be said of hysteresis in other fields such as structural
mechanics, tribology, magnetism and biology. It should be noted
that most of the existing analyses of dynamic systems with
hysteresis have their roots in earlier work
\cite{yakubovich1965,barabanov1979}. A combination of passivity
\cite{khalil2002}, Lyapunov \cite{lyapunov1892general}, and Popov
\cite{popov1961absolute} stability theories have been used
\cite{Pare2001} for analyzing the asymptotic stability and
stationary set of systems with hysteresis based on the LMI (linear
matrix inequality). Some strong assumptions, including loop
transformation with integral nonlinearity and smooth approximation
of discontinuities (for example for Preisach
\cite{preisach1935,KrasnoselProkr89,visintin1994} hysteresis
operator), have been requested in \cite{Pare2001}, while the
proposed stability theorem required formulating and solving an LMI
with certain additional conditions and constraints. Although more
recently the absolute stability of dynamic systems with hysteresis
operator has been addressed in \cite{ouyang2014}, here the
hysteresis was required to be the so-called Duhem operator
\cite{Oh2005}, and a certain matrix inequality, similar to KYP
(Kalman–-Yakubovich–-Popov) lemma \cite{khalil2002}, had to be
solved. Moreover, only examples of second-order systems have been
described. The related passivity property, which states that the
feedback connection of two passive systems is passive, has been
used for stability analysis of systems with hysteresis in
\cite{gorbet2001}. However, the passivity of hysteresis in
feedback has been established for the output rate only and not for
the output itself. Several analyses of dynamic behavior of systems
incorporating hysteresis have mainly been guided as
domain-specific, for example when studying friction where the
dissipative properties of hysteresis have been brought into focus,
see
\cite{Lazan68,dahl1976solid,koizumi1984study,AlBender2004,ruderman2017}.

The objective of this study is to contribute towards stability
analysis of systems with linear forward dynamics (of an arbitrary
order) and static hysteresis nonlinearity in the feedback channel.
While the dynamic hysteresis loops have also been described and
analyzed, e.g., in \cite{chua1971}, the rate-independent
hysteresis nonlinearity will be our main focus, for details see
\cite{Coleman1986}. It is worth noting that this type of
hysteresis, where rate-independent maps have a persistent memory
of the past after the transients have died out, is particularly
relevant for multiple natural and artificial systems, in other
words engineering systems. Furthermore it should be stressed that
our study focuses on the clockwise hysteresis, for which the
$(\dot{y},\xi)$ input-output system is passive, cf.
\cite{khalil2002} for passivity. This is despite the fact that
hysteresis with counter-clockwise input-output loops (sometimes
referred to as passive hysteresis \cite{Pare2001}), are usually
considered like Preisach operators, play-type operators and
phenomenological or physics-based magnetic hysteresis models. For
details see \cite{bertotti2006}. For analysis of the systems with
counter-clockwise input-output dynamics, we also refer to
\cite{Angeli2006}. While the counter-clockwise hysteresis in
systems is often in the forward channels of physical signals and
hence energy flow (e.g., classical B-H magnetic hysteresis), the
clockwise hysteresis usually appears in the feedback
interconnection. Examples include nonlinear friction forces, e.g.,
\cite{AlBender2004}, restoring forces in elasto-viscoplastic
structures, e.g., \cite{chaboche2008}, and vibro-impact in the
collision mechanisms \cite{hunt1975}. Several models for the
latter example can also be found in \cite{orlov1988}.

Before dealing with the main content of the paper, we first need
to formalize the hysteresis systems that we will focus on. These have
to fulfill two requirements: to be (i)
rate-independent and (ii) clockwise in the input-output (I/O)
sense. The first requirement means that the hysteresis output trajectories,
and consequently the energy conservation or dissipation, do not
depend on the input rate $\dot{y}$. This means a rate-independent
(also called static) hysteresis map is invariant to affine
transformations of time, i.e., $a+bt$ $\forall \; a\in
\mathbb{R},\, b \in \mathbb{R}^{+}$. The second requirement
implies that for any pair of the output trajectory segments which
are connected via an input reversal point, the forward segment for
which $\dot{y}>0$ always lies above the backward one for which
$\dot{y}<0$.

The rest of the paper is organized as follows. In Section
\ref{sec:2} we first discuss a dissipative clockwise hysteresis
map. This is to highlight its energy dissipation property upon the
input reversals. The linear dynamic system with hysteresis
feedback is introduced in Section \ref{sec:3} with the associated
equilibrium set. The main results in terms of the proposed loop
transformation and circle-based absolute stability are described
in Section \ref{sec:4}. Section \ref{sec:5} contains several
numerical examples which illustrate the stabilizing properties of
hysteresis in feedback and invariant sets of equilibria. Two
boundary cases of equilibria, lying on the coordinate axes of
hysteresis map, are demonstrated. Following this, another relevant
example of the feedback controlled double-mass harmonic oscillator
with hysteresis spring is described in detail. This includes
hysteresis with and without discontinuities and a hysteresis-free
case of memoryless nonlinear spring. Some general stabilizing
properties of hysteresis in the closed-loop are demonstrated and
discussed. Finally, the main conclusions are drawn in Section
\ref{sec:6}.

\section{Dissipative hysteresis map}
\label{sec:2}

Considering a static hysteresis map with the input $y$ and output
$\xi$, which represent ''energy pair`` of a physical system, one
can introduce the supply rate $w = \dot{y}(t) \cdot \xi(t)$. The
energy flowing into the system, I/O hysteresis in our
case, is obtained as
$$
\int_{t_1}^{t_2} w \mathrm{d}t.
$$
Then, for a non-negative real function $V(z): Z \rightarrow
\mathbb{R}^{+}$ of the system state $z$, called the storage
function, the system is said to be \emph{dissipative}, cf.
\cite{willems2007}, when satisfying
\begin{equation}
V(z(t_2)) - V(z(t_1)) \leq \int_{t_1}^{t_2} w \mathrm{d}t
\label{eq:2:1:0}
\end{equation}
for all $(y,\xi,z)$ and $t_2 \geq t_1$. In other words, for
all causal time trajectories, a dissipative system cannot store
more energy than is supplied to it, while the difference represents
the energy dissipated by the system. It is worth noting that the
scalar function $V(\cdot)$ is often also called the energy
function, or more specifically the Lyapunov function, since it reflects
the instantaneous energy level of the lumped system supplied
through $y$.

\begin{figure}[!h]
\centering
\includegraphics[width=0.5\columnwidth]{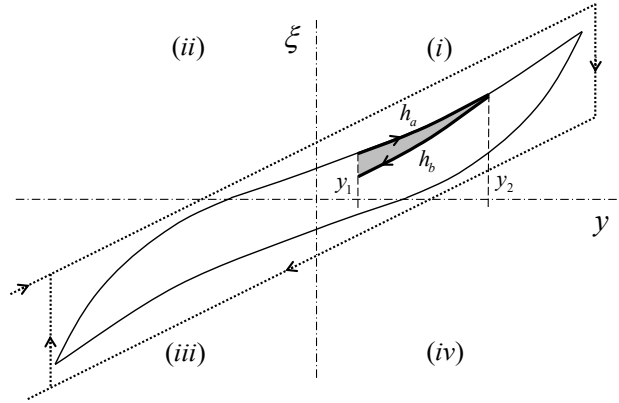}
\caption{Hysteresis loop $\xi(y)$ with input reversals.}
\label{fig:2:1}
\end{figure}

Next, we will consider the energy level of hysteresis system
$\xi(y)$, which is indirectly analogous to a hysteresis spring
which has a displacement-to-force ''energy pair``. Any changes in
energy level and corresponding energy storage, can be considered for
two arbitrary points $y_{2} > y_{1}$, assumed as exemplary and
depicted in Figure \ref{fig:2:1} without loss of generality. Here
$y_{1}$ represents an initial state and $y_{2}$ represents the
next reversal state in the input sequence. Recall that the total
energy state of hysteresis system can then be written as
\begin{equation}\label{eq:2:1:1}
E(t) = \int_{Y(t)} \xi(y) \mathrm{d} y + E(0),
\end{equation}
where $Y(t)$ is the entire hysteresis input path with $Y(0)=y_{1}$
and $E(0)$ is the initial value; for the sake of simplicity we
assume $E(0)=0$. Note that no kinetic-type energy should be taken
into account since the input-output hysteresis map is
rate-independent \cite{visintin1994,BrokateSprekels96}, i.e.,
$E(t)$ is independent of the input rate.
For a monotonically increasing trajectory $Y(t) \in [y_{1},
y_{2}]$ it can be seen that the change in energy level is given by
\begin{equation}\label{eq:2:1:2}
E(y_{2}) - E(y_{1})= \int^{y_{2}}_{y_{1}} \xi(y) \mathrm{d}y.
\end{equation}

First, one can show that if $\xi(y)$ has no hysteresis effect,
i.e. no loop, then the forward and backward trajectories coincide
with each other. Therefore, for the difference in energy levels at
both points, one can write
\begin{eqnarray}
\label{eq:2:1:3}
  \Delta E_{2-1} &=& \int^{y_{2}}_{y_{1}} \xi_{a}(y) \mathrm{d}y = -\int^{y_{1}}_{y_{2}} \xi_{b}(y) \mathrm{d}y,
  \\[0.2mm]
  \Delta E_{1-2} &=& \int^{y_{1}}_{y_{2}} \xi_{b}(y) \mathrm{d}y = -\int^{y_{2}}_{y_{1}} \xi_{a}(y) \mathrm{d}y.
\label{eq:2:1:4}
\end{eqnarray}
Obviously, the total energy change during one
$y_{1}$-$y_{2}$-$y_{1}$ cycle is $\Delta E_{2-1}+\Delta
E_{1-2}=0$. That means the system behaves as lossless, i.e.,
without hysteresis-related damping.

Next, we consider an actual hysteresis branching after an input
reversal, as shown in Figure \ref{fig:2:1}, so that the
corresponding energy change results in
\begin{equation}\label{eq:2:1:5}
\Delta E = \Delta E_{2-1} + \Delta E_{1-2} = \int^{y_{2}}_{y_{1}}
\xi_{a}(y) \mathrm{d} y - \int^{y_{1}}_{y_{2}} \xi_{b}(y)
\mathrm{d} y \neq 0.
\end{equation}
The first summand in the above equation constitutes the potential
energy which has been stored when moving from $y_{1}$ to $y_{2}$,
while the second summand corresponds to the potential energy which
has been released on the way back. The positive difference $\Delta E
> 0$ represents the energy losses due to the hysteresis. Obviously,
the amount of loss corresponds to the area of hysteresis loop,
gray-shaded in Figure \ref{fig:2:1}. It is worth noting that the
loop does not necessarily need to be closed. As the input proceeds after
reversal, the growing (gray-shaded) area reflects the increment
in energy losses during each cycle. That is, each flip in the sign
of the input rate results in energy dissipation, provided the
forward and backward paths diverse from each other. This allows
us to introduce the following
\begin{equation}\label{eq:2:1:6}
\xi(y) = g(y) + h \, \mathrm{sign} (\dot{y})
\end{equation}
input-output hysteresis map (equally drawn in Figure \ref{fig:2:1}
by the dotted-line), as the upper boundary of energy dissipation
after input reversals. Here $h > 0$ is a parametrization constant
and $g(\cdot)$ is a static (generally nonlinear) map satisfying
the sector conditions, cf. \eqref{eq:4:1:2} below. In the
following, we will assume a linear relationship $g(y) = \gamma \,
y$ with the slope constant $\gamma > 0$, this is for the sake of
simplicity and without loss of generality. Furthermore, it is
worth noting that for $h \rightarrow 0$ the input-output map
\eqref{eq:2:1:6} approaches the lower boundary case of energy
dissipation $\Delta E \rightarrow 0$, with a corresponding vanishing of
hysteresis. It seems appropriate to assume that all clockwise
hysteresis maps which lie between the lower and upper boundary
cases, as defined above, will behave as dissipative upon the input
cycles. Moreover, this is independent of the shape of the hysteresis
loops.

For the input-output hysteresis map the following storage
function, fully analogous to the generalized energy function of
hysteresis spring, can be introduced
\begin{equation}
V(y) = \frac{1}{2} \gamma \, y^2 - \int_{Y} \xi(y)\mathrm{d}y.
\label{eq:2:1:7}
\end{equation}
Note that while the first term of \eqref{eq:2:1:7} describes the
potential energy which is stored and therefore can be
recuperated, the second term constitutes the energy dissipation
over the path $Y$ traversed within the hysteresis loop. It is
worth emphasizing that this class of Lyapunov, corresponding to
energy functionals, has already been introduced in
\cite{yakubovich1965}, for analyzing systems with a hysteresis
nonlinearity. Taking time derivative of the storage function
\eqref{eq:2:1:7}, for the input-output map \eqref{eq:2:1:6}
with $g=\gamma \, y$, one obtains
\begin{equation}
\dot{V} = \gamma \, y \, \dot{y} - \gamma \, y \, \dot{y} - h \,
\mathrm{sign}(\dot{y}) \, \dot{y} = - h |\dot{y}|.
\label{eq:2:1:8}
\end{equation}
From the above, one can see that the time derivative of the
storage (or Lyapunov) function is always negative definite if
$\dot{y} \neq 0$. Only for $h=0$ does the system behave as lossless,
which means that no hysteresis loops arise after input reversals.
Otherwise, any input changes would provoke an energy dissipation at
a rate proportional to the input rate. Therefore, a
clockwise hysteresis, enclosed by any system dynamics, will act as
a damping element on the input cycles. However, no exponential
damping rate, as otherwise conventional for linear system
dynamics, can be associated with hysteresis. It appears natural
since no Laplace transform corresponding to poles allocation is
available for analysis of hysteresis behavior. Below, we
address the stability of linear dynamic systems incorporating
hysteresis in feedback. This is in the spirit of the Lure's type systems
\cite{lur1944theory} which have been thoroughly investigated and formalized for
nonlinearities without memory.

\section{Feedback system with hysteresis}
\label{sec:3}

The considered feedback system
\begin{equation}
\Phi = \left\{%
\begin{array}{ll}
    \dot{x} = Ax + Bu, \\[0.2cm]
    y = C{^T}x,  \\[0.2cm]
    - u = \xi[y,t,\xi_0].
\end{array}%
\right. \label{eq:3:1:1}
\end{equation}
largely consists of a linear dynamics system, given in the
state-space notation and hysteresis nonlinearity $\xi$ connected
in negative feedback. The linear dynamics system is described by
the $n \times n$ system matrix $A$ and $n \times 1$ input and
output coupling vectors $B$ and $C$, respectively. The system
output of interest is $y$, while $x$ is a $n \times 1$ vector of
the system states. The initial hysteresis state is denoted by
$\xi_0$. The time argument $t$ is omitted for the sake of
simplicity, unless in $\xi$, and that is for explicitly stressing
the memory properties of the hysteresis operator. That means the
output at $t$ depends from its previous values on the time
interval $[0, t)$. Generally speaking, $\Phi$ is what we will
understand under the dynamic system with hysteresis feedback, in
equivalence to the Lure's systems \cite{lur1944theory} which have
memoryless nonlinearities. Note that although $u$ is often assumed
to be a control variable, it can equally appear as some coupling
quantity for hysteresis nonlinearity in feedback.

Let us now define the equilibrium state, corresponding to the
equilibrium set due to the set-valued hysteresis nonlinearity.
It was provided in an elegant way in \cite{barabanov1979}.
Introducing the corresponding transfer function
\begin{equation}
G(s) = C^{T}(sI - A)^{-1}B, \label{eq:3:1:2}
\end{equation}
which is a dynamic map from $u$ to $y$, with $s$ as the Laplace
variable, one can write a steady-state equation
\begin{equation}
y + G(0)\xi = 0 \label{eq:3:1:3}
\end{equation}
for hysteresis nonlinearity acting in feedback. This one
represents a line in the $(y,\xi)$-coordinates, which cuts the
hysteresis loop at the corresponding points $\{(y_0,\xi_0)\}$. It
is obvious that the steady-state of the dynamic system $\Phi$, cf.
\eqref{eq:3:1:1}, is given by
\begin{equation}
x_0 = A^{-1}B \xi_0, \label{eq:3:1:4}
\end{equation}
while $x(t)=x_0$ constitutes all steady-state solutions of the
closed-loop system \eqref{eq:3:1:1}. It implies that the
equilibrium set $E=\{x_0\}$ is built up from those $(x_0,\xi_0)$
solutions for which \eqref{eq:3:1:3} and \eqref{eq:3:1:4} hold,
and that they are independent of the initial $(\xi(0),x(0))$. Recall that
due to a multi-valued hysteresis map, the initial $\xi(0)$ should
be explicitly taken into account, along with $x(0)$ of the dynamic
system.

\section{Loop transformation and absolute stability}
\label{sec:4}

\begin{figure}[!h]
\centering
\includegraphics[width=0.4\columnwidth]{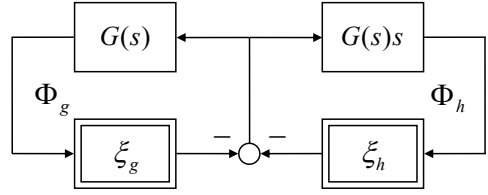}
\caption{Transformed loop with both nonlinearities}
\label{fig:4:1:1}
\end{figure}

Considering the boundary case, cf. with \eqref{eq:2:1:6}, of
hysteresis nonlinearity in feedback, one obtains the output of
$\Phi$ as a Laplace transformed superposition
\begin{equation}
y(s) = -G(s) \, \xi_g \left[ y(s) \right] - G(s) \, \xi_h \left[
y(s) \, s \right] \label{eq:4:1:1}
\end{equation}
of both input parts in the loop. Here $\xi_g [ \cdot ]$ maps the
static part of feedback nonlinearity, not necessarily limited to
the linear $g$-gain, while $\xi_h [ \cdot ] = h \, \mathrm{sign} [
\cdot ]$ is the sign-nonlinearity with output rate in argument. By
means of the block diagram algebra, with corresponding loop
transformations, the closed-loop \eqref{eq:4:1:1} can be brought
into the structure as shown in Figure \ref{fig:4:1:1}.
Both closed loops, coupled in the way of superimposed feedback,
are denoted by $\Phi_g$ and $\Phi_h$ respectively.
One can recognize that while both linear
transfer functions are subject to the common input $u$, the
outputs of both nonlinearities act as external disturbances, each
one for the opposite loop, i.e., $u_g=\xi_g$ for $\Phi_h$ and
$u_h=\xi_h$ for $\Phi_g$. This allows us to investigate
the input-output stability of $\Phi_g$ and $\Phi_h$ separately
from each other, as shown in the following.

\begin{figure}[!h]
\centering
\includegraphics[width=0.5\columnwidth]{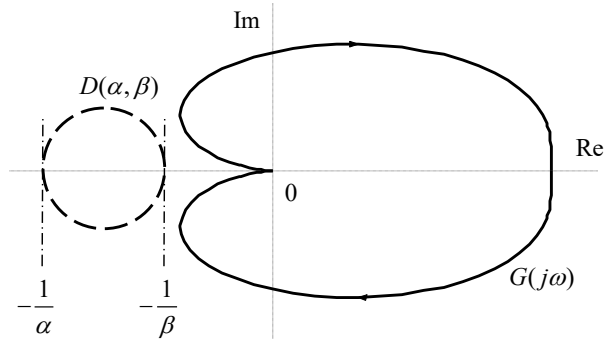}
\caption{Graphical interpretation of circle criterion}
\label{fig:4:1:2}
\end{figure}

The widely accepted circle criterion
\cite{sandberg1964,zames1966input,brockett1967frequency}, which is
closely related to and can be derived from the Popov theorem
\cite{popov1961absolute} of absolute stability of the Lure's type
systems \cite{lur1944theory}, enables analysis of the dynamical
systems $\Phi$. The latter includes $\xi$ which should satisfy the
$[\alpha, \beta]$ sector condition, cf. e.g., \cite{khalil2002},
\begin{equation}
\alpha \, y^2 \leq \xi(y)\, y \leq \beta \, y^2.
 \label{eq:4:1:2}
\end{equation}
Not that the $[0, \infty)$ sector, as a particular case, is also
included. The circle criterion says that system matrix $A$ has
no eigenvalues on the $j\omega$ axis and $v$ eigenvalues strictly
in the right half-plane. The Nyquist locus of $G(j\omega)$ does
not enter the critical disk $D(\alpha, \beta)$ and encircles it
$v$ times counter-clockwise, that for $0 \leq \alpha \leq \beta$,
cf. \cite{SlotLi90,sastry2013nonlinear}. Recall that the open disk
$D(\alpha, \beta)$ lies in the complex half-plane $\mathrm{Re} [s]
\leq 0$ passing through $-\alpha^{-1}$ and $-\beta^{-1}$ as
 shown in Figure \ref{fig:4:1:2}.
Also note that the critical disk shrinks to the conventional
critical point as the gain limits $\alpha$ and $\beta$ approach
each other \cite{brockett1967frequency}, while for nonlinearities
lying in the sector $[0, \beta]$ the critical disk ''blows up``
into the half-plane $\mathrm{Re}[s] \leq -\beta^{-1}$. The latter
allows reducing the general circle criterion to
\begin{equation}
\mathrm{Re} \left[1 + \beta G(j\omega) \right] > 0, \quad \forall
\, \omega,
\label{eq:4:1:3}
\end{equation}
cf. \cite{sastry2013nonlinear}, that ensures the origin of $\Phi$
to be asymptotically stable, while the above stability of $A$ is
in effect. Note that for the case when inequality \eqref{eq:4:1:3}
contains $\geq 0$, meaning the Nyquist locus $G(j\omega)$ is
admitted to touch the vertical axis $-\beta^{-1}$, only a bounded
output stability can be guaranteed, cf.
\cite{brockett1967frequency}.

In the following, we will make a restrictive assumption of a
stable linear system dynamics, thus requiring $v=0$. Therefore,
non-entering of the critical disk remains the single sufficient
(but not necessary) condition to be fulfilled by the circle
criterion. With regard to the transformed closed-loop of $\Phi$,
cf. Figure \ref{fig:4:1:1}, we request that both $\Phi_g$ and
$\Phi_h$ satisfy the above circle criterion. This is necessary for
the entire system $\Phi$ is asymptotically stable. For $\Phi_g$,
the situation is standard and for the given $\xi_g \in [0,
\beta]$, it is straightforward to evaluated the fulfillment of
circle criterion. For $\Phi_h$, which matters in the hysteresis
loop, one can show that $\beta \rightarrow \infty$ while
$\alpha=\beta$ due to the sign-operator which switches immediately
upon zero crossing of the input of $\xi_h$. Here we stress that
the sign nonlinearity lies completely in the first and third
quadrants, thus complying with the sector condition of
nonlinearity in feedback. The latter is of relevance to avoid
violating the formulation of Lure's problem, as in
\eqref{eq:3:1:1} cf.
\cite{lur1944theory,khalil2002,sastry2013nonlinear}, for which the
circle criteria have been originally elaborated. The above sector
condition yields that for $\Phi_h$, the critical disk
$D(\infty,\infty)=0$ shrinks to the single point, so that the
Nyquist locus of $j\omega G(j\omega)$ is requested to not enter,
or correspondingly encircle, the coordinate origin $\forall \;
|\omega|>0$.

It is worth emphasizing that the circle criterion specified above
for $\xi$-type hysteresis nonlinearities is sufficient to prove
the global asymptotic stability of $\Phi$. However, based on the
above, nothing can be said about the possible stabilizing
properties of hysteresis in feedback. This means that for the case
when $G(s)$ has unstable poles lying in $\mathbb{C}_{+}$.
Notwithstanding, a numerical example of stabilizing hysteresis
properties will be provided and commented on in Section
\ref{sec:5:2}.

\section{Numerical examples}
\label{sec:5}

\subsection{Boundary cases with simple dynamics}
\label{sec:5:1}

We look at two boundary cases for which the intersection line of
hysteresis nonlinearity \eqref{eq:2:1:6} coincides with one of the
$(y,\xi)$ axes. Recall that \eqref{eq:2:1:6} captures a limiting
shape of the clockwise hysteresis that features the maximal energy
dissipation upon changes in the input direction. Here and for the
rest of subsection, the unity $g,h$ coefficients are assumed for
the sake of simplicity. Recall that hysteresis is in negative
feedback with the linear dynamics, as in $\Phi$ defined in Section
\ref{sec:3}.

\begin{figure}[!h]
\centering
\includegraphics[width=0.3\columnwidth]{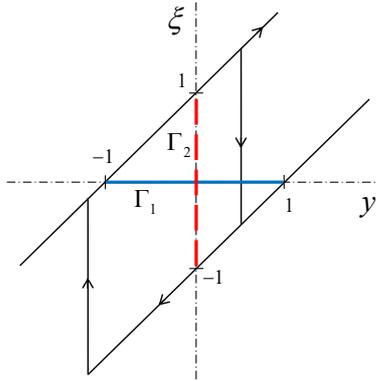}
\caption{Graphical representation of equilibrium sets within
hysteresis nonlinearity} \label{fig:5:1:1}
\end{figure}

\begin{figure}[!h]
\centering
\includegraphics[width=0.45\columnwidth]{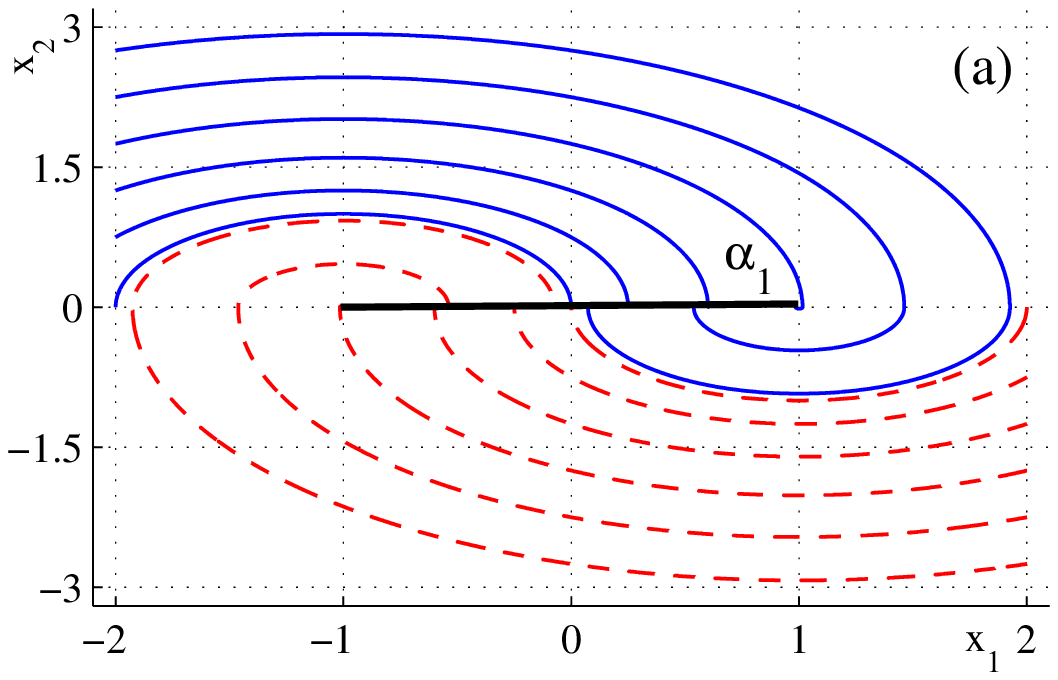}
\hspace{0.1cm}
\includegraphics[width=0.45\columnwidth]{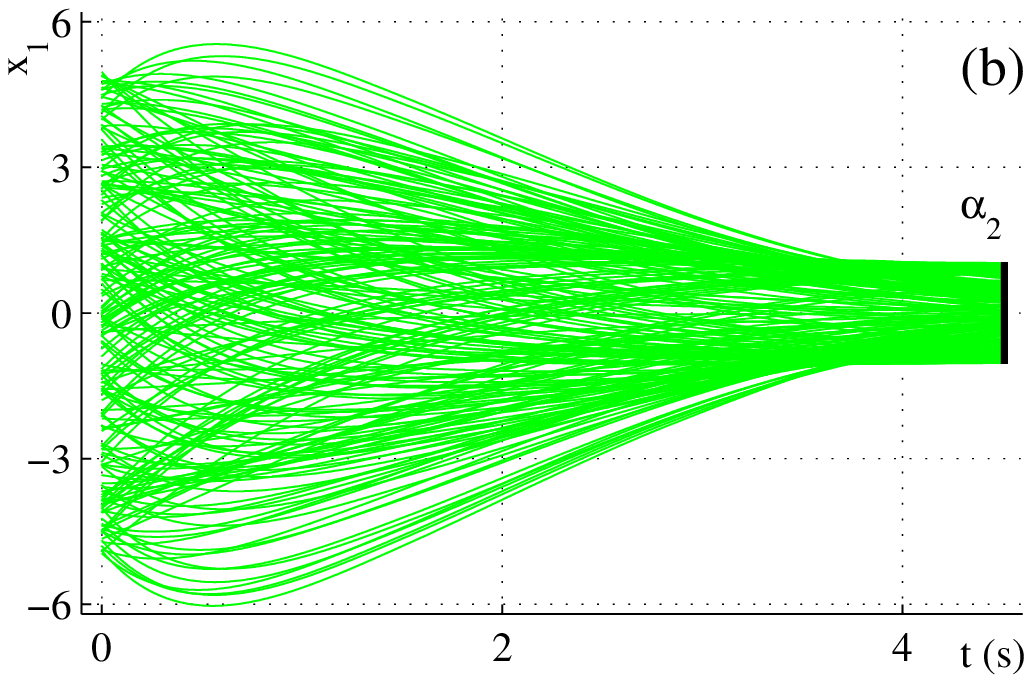}
\caption{Trajectories with attraction to the invariant sets, for
$G_1$ in (a) and $G_2$ in (b)} \label{fig:5:1:2}
\end{figure}

First, assume a double integrator system, with the state vector
$x=(x_1,x_2)^T$, so that
\begin{equation}\label{eq:5:1:1}
A_1 = \left(%
\begin{array}{cc}
  0 & 1 \\
  0 & 0 \\
\end{array}%
\right), \quad
B_1 = \left(%
\begin{array}{c}
  0 \\
  1 \\
\end{array}%
\right), \quad
C_1^{T} = \left(%
\begin{array}{cc}
  1 & 0 \\
\end{array}%
\right).
\end{equation}
It is evident that the corresponding transfer function
$G_1(0)\rightarrow \infty$ and the $\Gamma_1$-line rests on the
$y$-axis cf. with \cite{barabanov1979}, while it cuts the
hysteresis loop as schematically shown in Figure \ref{fig:5:1:1}.
Since $A_1$ is singular, i.e., not invertible, no equilibrium state
can be found by $x_0(t) = A_1^{-1} B_1 \xi_0$ for $y_0 + G_1(0)
\xi_0 = 0$ belonging to $\Gamma_1$. At the same time,
\begin{equation}\label{eq:5:1:2}
S_1 = \{ x \in \mathbb{R}^2 \; |  \; x_1 \in \Gamma_1, \: x_2 = 0
\}
\end{equation}
constitutes an invariant (or stationary) set, in which the
trajectories are independent of the initial state $x(0)$.
Each point of $S_1$ represents a steady-state solution of the
hysteresis closed-loop \eqref{eq:3:1:1} with \eqref{eq:2:1:6} and
\eqref{eq:5:1:1}, while $\alpha_1 = \{-1 \leq x_1 \leq 1, x_2=0
\}$ acts as a global attractor in the state-space, as
shown in Figure \ref{fig:5:1:2} (a).
Note that the double integrator dynamics does not fulfill the
circle criterion, cf. Section \ref{sec:4}, and that for the static
(memoryless) part $\xi_g$ of feedback nonlinearity.

Next, we assume a second-order system
\begin{equation}\label{eq:5:1:3}
A_2 = \left(%
\begin{array}{cc}
  0 & 1 \\
  -1 & -1 \\
\end{array}%
\right), \quad
B_2 = \left(%
\begin{array}{c}
  0 \\
  1 \\
\end{array}%
\right), \quad
C_2^{T} = \left(%
\begin{array}{cc}
  0 & 1 \\
\end{array}%
\right).
\end{equation}
for which the transfer function $G_2(0) = 0$. From $y_0 + G_2(0)
\xi_0 = 0$ one obtains the vertical $\Gamma_2$-line which cuts the
hysteresis at $y_0 = 0$, as equally visualized in Figure
\ref{fig:5:1:1}. Here one should stress that the
$(y,\xi)$-coordinates are different for $G_1$ and $G_2$
input-output dynamics, while the hysteresis nonlinearity in both
cases is the same, given by \eqref{eq:2:1:6}. According to
\eqref{eq:2:1:6}, the $\xi_0 \in [-1, 1]$ at $y_0 = 0$, while the
sign-dependency is governed by
$\mathrm{sign}\bigl(\dot{y}_0(t^{-})\bigr)$, i.e., instantly
before reaching the output steady-state $y_0$. From the
equilibrium $x_0(t) = A_2^{-1} \, B_2 \, \xi_0$, for $\xi_0 = \pm
1$, one obtains the lower and upper boundaries $x_0 = \bigl\{
\{-1,1\}, 0 \bigr\}$ of the invariant set
\begin{equation}\label{eq:5:1:4}
S_2 = \{ x \in \mathbb{R}^2 \; | \; x_1 \in [-1, 1], \: x_2 = 0
\}.
\end{equation}
Here again, each point of $S_2$ represents a steady-state solution
of the hysteresis closed-loop \eqref{eq:3:1:1} with
\eqref{eq:2:1:6} and \eqref{eq:5:1:3}, while $\alpha_2 = \{-1 \leq
x_1 \leq 1, x_2=0 \}$ acts as a global attractor of the state
trajectories. A set of $x_1(t)$ trajectories, converging to
$\alpha_2$, are shown in Figure \ref{fig:5:1:2} (b) for a large
number of random $x(0)$ initializations. Referring to the circle
criterion, cf. Section \ref{sec:4}, one should notice that the
Nyquist locus of $G_2(s)s$ is touching the critical disk for
$\omega = 0$. Recall that the critical disk for the
$\mathrm{sign}$-nonlinearity collapses to the single critical
point, thus approaching zero from the left for $\beta \rightarrow
\infty$. In this case, $\Phi$ is stable in the sense that all sets
of initial conditions lead to a bounded $y$ as $t \rightarrow
\infty$ \cite{brockett1967frequency}. Moreover, the Nyquist locus
enters the critical point only in the stationary case
($\omega=0$), that cancels stability issues for a $\Phi$ system
which has free differentiators in the loop.

\subsection{Feedback controlled double-mass harmonic oscillator}
\label{sec:5:2}

\begin{figure}[!h]
\centering
\includegraphics[width=0.35\columnwidth]{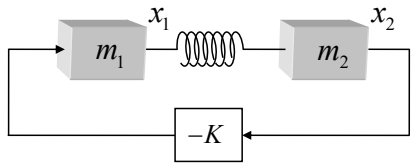}
\caption{Closed-loop harmonic oscillator with nonlinear spring}
\label{fig:5:2:1}
\end{figure}

Consider a feedback controlled two-mass harmonic oscillator, for
which the connecting spring features the nonlinear stiffness with
hysteresis. The free body diagram, shown in Figure
\ref{fig:5:2:1}, visualizes the corresponding closed-loop dynamic
system. The setup represents a class of common engineering
problems where the inertial elements are interconnected by an
energy storage which is coupled with (structural) energy
dissipation that takes place during harmonic oscillations.
Examples for this can be found in the machine elements
\cite{Dhaouadi2008,Ruderman2015}, piezoelectric materials
\cite{Goldfarb1997,damjanovic2006}, micro-frictional and generally
tribological systems \cite{Socoliuc2004,AlBender2004}, structures
with elasto-plasticity \cite{Gerstmayr2003,chaboche2008} and
others. An open-loop system is inherently
passive and therefore remains stable even if a harmonic
oscillator without auxiliary damping can reveal nontrivial
transient trajectories in the presence of the hysteresis. However, in
order to explicitly address the stability of the closed-loop
dynamics with hysteresis, a negative feedback of $x_2$-state is
incorporated. This allows for a direct analogy, for instance to
an output feedback controller, where $K > 0$ is the control gain
of an appropriate choice. Note that despite mimicking an output
feedback control structure, we deliberately omit to insert any
control damping. This is intentional since our goal is to
demonstrate the stability and possible stabilizing properties of
hysteresis in the loop. For the assumed unity masses, the
closed-loop system \eqref{eq:3:1:1} is given by
\begin{equation}\label{eq:5:2:1}
A_3 = \left(%
\begin{array}{cccc}
  0 & 0 & 1 & 0 \\
  0 & 0 & 0 & 1 \\
  -g & g-K & 0 & 0 \\
  g & -g & 0 & 0.01
\end{array}%
\right), \quad
B_3 = \left(%
\begin{array}{c}
  0 \\
  0 \\
  -1 \\
  1
\end{array}%
\right), \quad
C_3^{T} = \left(%
\begin{array}{cccc}
  1 & -1 & 0 & 0 \\
\end{array}%
\right),
\end{equation}
while for the static map $\xi$ we will consider both a memoryless
stiffness and stiffness with hysteresis. These are addressed
separately in the following subsections. The state vector is
$x=(x_1, x_2, x_3, x_4)^T$, while non-zero initial conditions
$x(0)=(1, 0, 0, 0)^T$ will be assumed for the solutions shown
below. Note that the linear stiffness of connecting spring,
assigned to be $g=100$, is already captured in the system matrix
$A_3$. Further we note that an auxiliary (low-valued) viscous
damping term $0.01 \times x_4$ is added to the second mass, in
order to provide a better separation between two pairs of the
conjugate-complex poles of $A_3$. This is also to numerically
stabilize computation of the solutions since we are addressing two
boundary cases in the vicinity of the stability margin of the
closed-loop system.

\subsubsection{Memoryless stiffness} \label{sec:5:2:1}

\begin{figure}[!h]
\centering
\includegraphics[width=0.4\columnwidth]{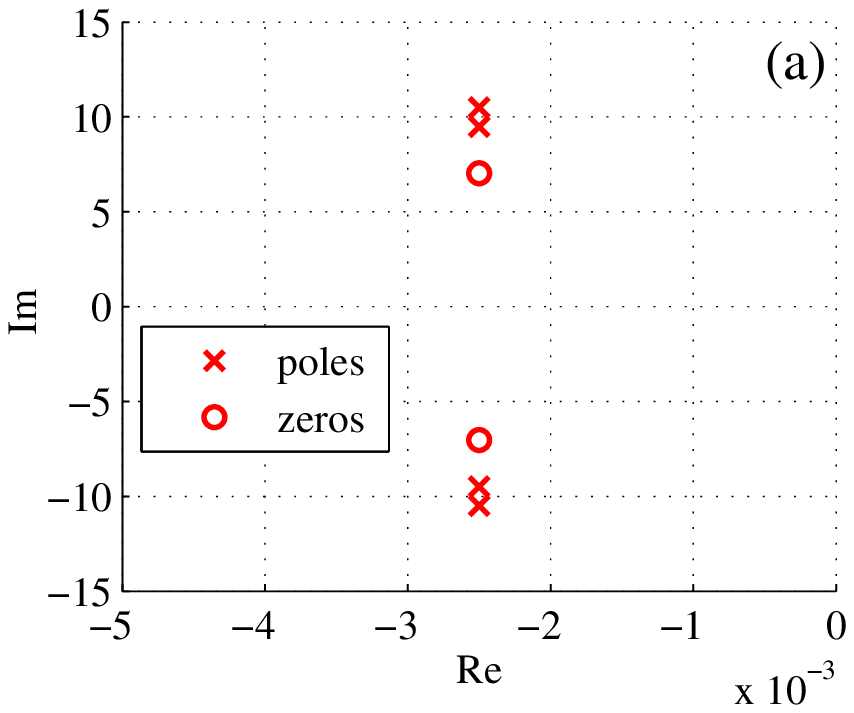} \hspace{0.1cm}
\includegraphics[width=0.4\columnwidth]{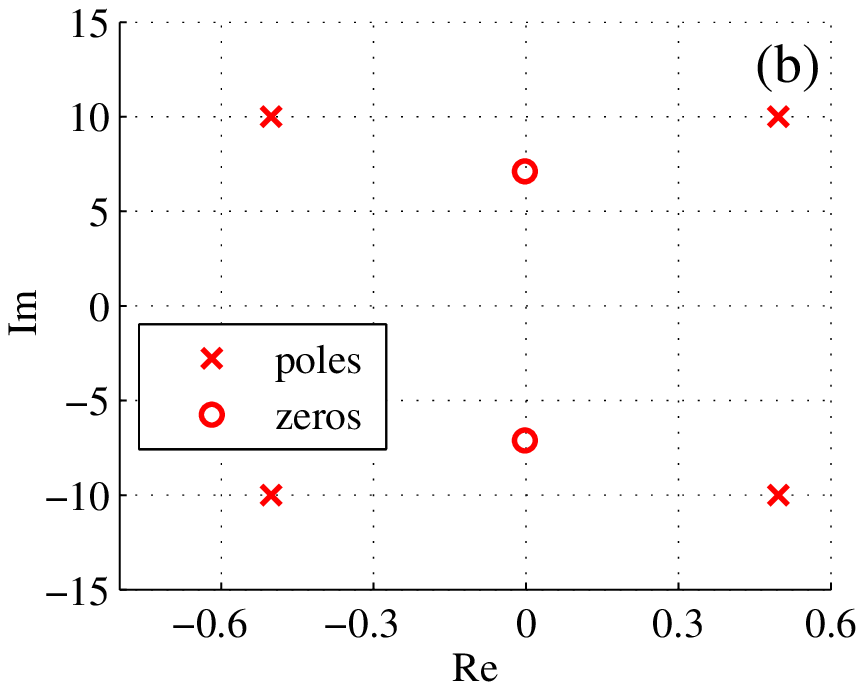}
\caption{Poles and zeros for boundary-stable $K=99$ (a) and
boundary-unstable $K=101$ (b) feedback} \label{fig:5:2:2}
\end{figure}

\begin{figure}[!h]
\centering
\includegraphics[width=0.4\columnwidth]{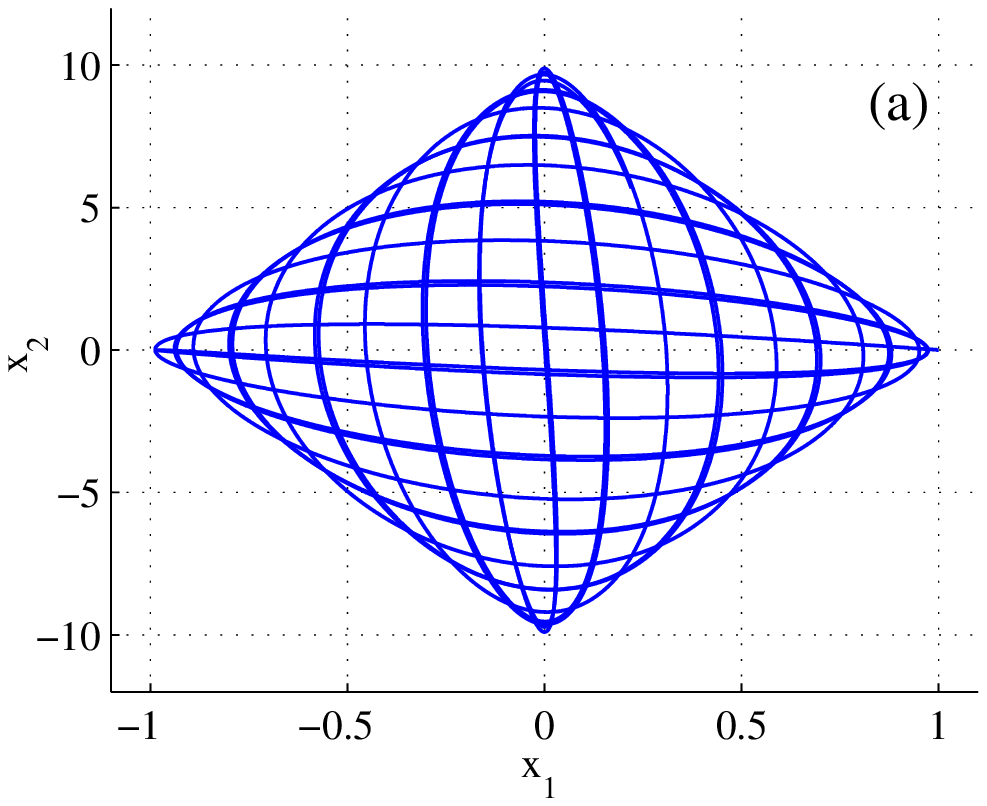} \hspace{0.1cm}
\includegraphics[width=0.4\columnwidth]{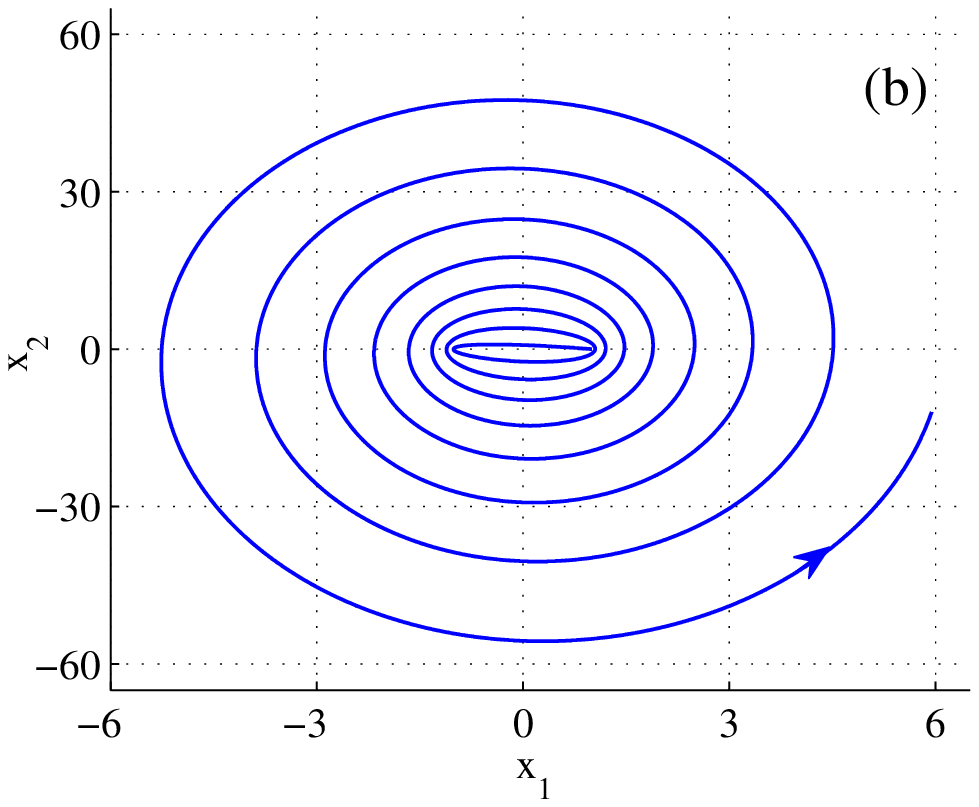}
\caption{Phase portrait of the $(x_1,x_2)$-states, once for
boundary-stable $K=99$ (a) and once for boundary-unstable $K=101$
(b) feedback gain} \label{fig:5:2:3}
\end{figure}

First, we consider a linear memoryless stiffness of the connection
spring so that $\xi = 0$. In this case an oscillating
behavior of the system will depend on assignment of
the feedback gain. Analyzing the system matrix $A_3$, one can show
that the closed-loop system becomes marginally stable once $K=g$.
For a stable and unstable system response, we
define a boundary-stable and boundary-unstable feedback gain
$K=99$ and $K=101$, respectively. The associated poles-zeros
configuration for both cases are shown in Figure \ref{fig:5:2:2}
(a) and (b). One can see that a relatively small variation of the
feedback gain, i.e., by some 2 \%, shifts one boundary-stable
conjugate complex pole pair into the right half-plane of the
$s$-domain. This results in an unstable configuration of the
system poles and,
as implication, the overall closed-loop dynamic system becomes
unstable, as can be seen from the phase portrait depicted in
Figure \ref{fig:5:2:3} (b). Opposite, the phase portrait of the
same $(x_1,x_2)$-states is shown in Figure \ref{fig:5:2:3} (a) for
the boundary-stable feedback gain.

\subsubsection{Hysteresis stiffness}
\label{sec:5:2:2}

Next, we give an example of the stabilizing property of hysteresis
in the loop, when keeping the boundary-unstable $K=101$ feedback
gain, as addressed above. The linear dynamics of the system $\Phi$
remains the same as in Section \ref{sec:5:2:1}, while the single
modification is made in the nonlinear part of \eqref{eq:2:1:6}.
We are assuming the sign-dependent hysteresis term with
$h=50$, further denoted as \emph{sign-hysteresis}. In this case
the feedback nonlinearity coincides exactly with \eqref{eq:2:1:6}.
Another, instead of $h \, \mathrm{sign}(\dot{y})$, we then assume
a simple single Prandtl-Ishlinskii (PI) stop-type operator
\cite{KrasnoselProkr89,Krejci96}, which represents one rheological
element, further denoted by \emph{PI-hysteresis}. Recall that
PI-hysteresis produces an additional slope in the $(y,\xi)$
coordinates, thus avoiding the sign-dependent step-wise
transitions. The single rheological element is parameterized by
two constants, one giving the output saturation level (analogous
to $h$ for the sign-hysteresis), and one determining the slope
after the input reversals. For details see \cite{ruderman2017}.

\begin{figure}[!h]
\centering
\includegraphics[width=0.4\columnwidth]{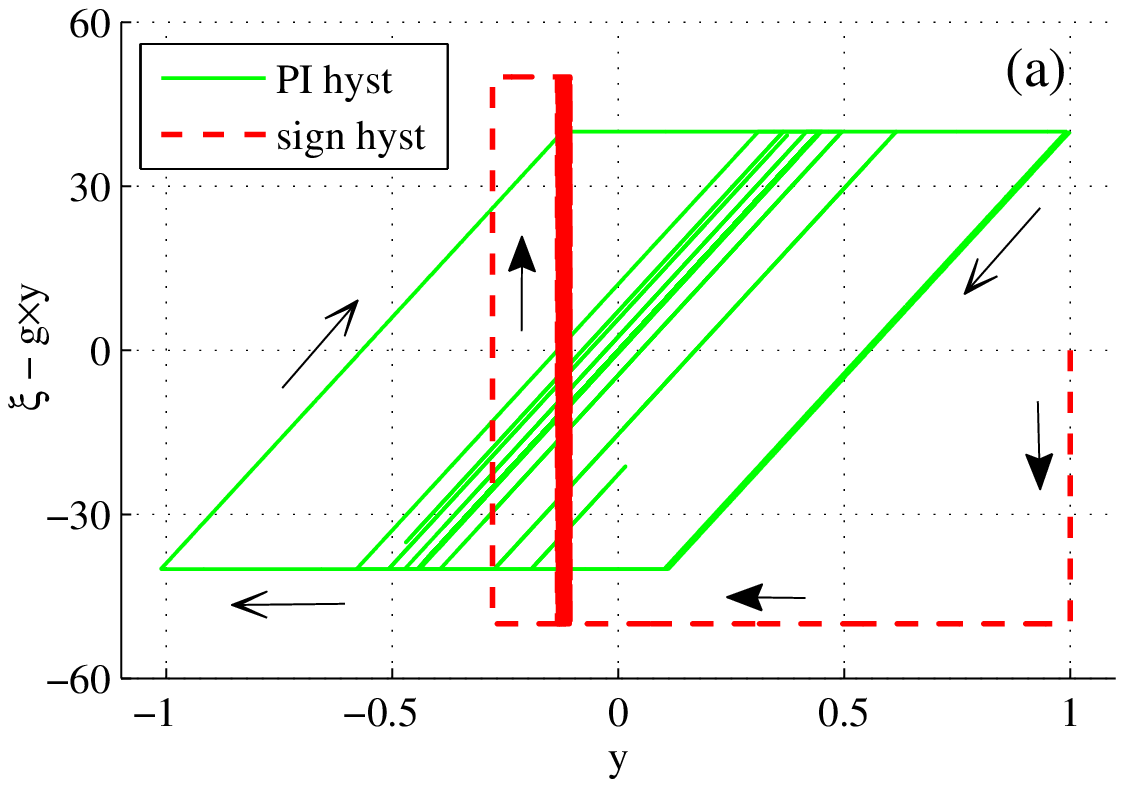} \hspace{0.1cm}
\includegraphics[width=0.4\columnwidth]{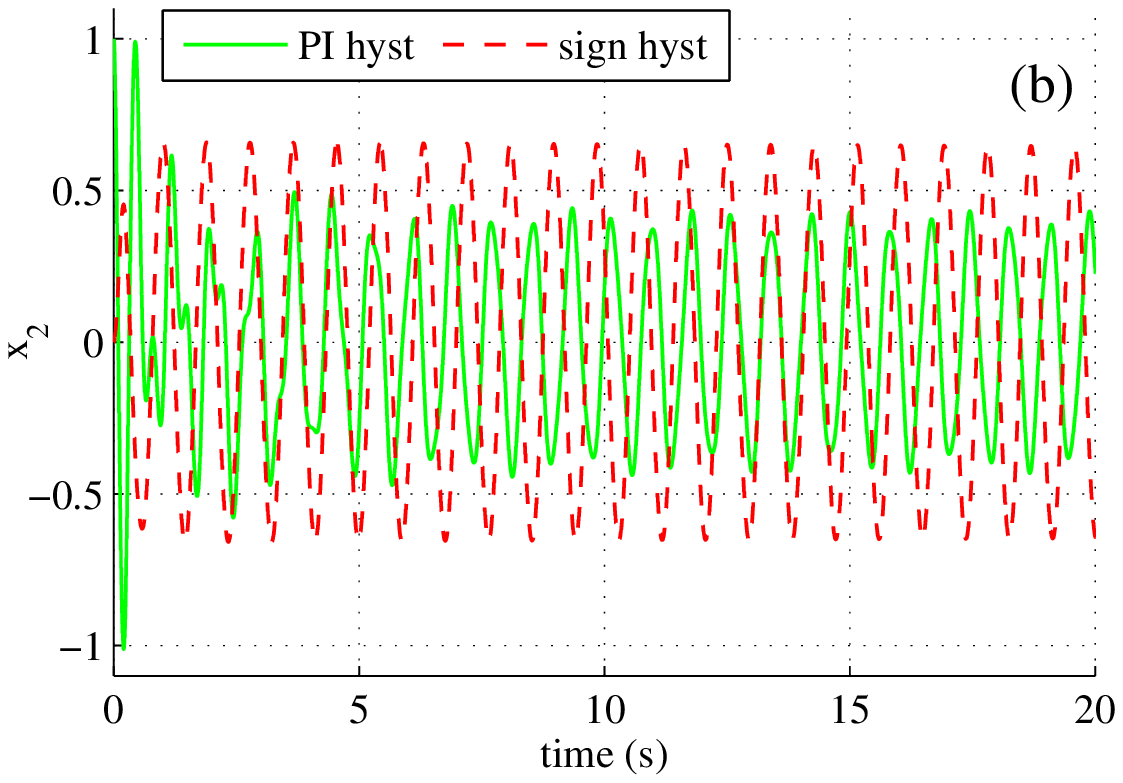}
\caption{Response of feedback nonlinearity \eqref{eq:2:1:1}
(plotted without linear stiffness part) for both PI- and
sign-hysteresis (a), and corresponding $x_2$-state response (b)}
\label{fig:5:2:4}
\end{figure}

For the closed-loop response, with the same initial conditions as
in Section \ref{sec:5:2:1}, the input-output map of both
nonlinearities are shown in Figure \ref{fig:5:2:4} (a),
without incorporating the linear stiffness part (already captured
in $A_3$). The sign-hysteresis converges to a final state, while
the numerical integration at discontinuities produces a (minor)
spurious drift, see the red vertical bar in Figure \ref{fig:5:2:4}
(a). The same can be said of the PI-hysteresis trajectory while
multiple reversal slopes still remain inside the surrounding
major loop. The corresponding $x_2$-state trajectories, shown in
Figure \ref{fig:5:2:4} (b), indicate the appearance of a
stable limit cycle in both cases, while the second superimposed
harmonic, obviously related to additional stiffness of
PI-hysteresis, is also visible during initial transients.

Note that in both the above examples, inclusion of
the hysteresis provides a stabilizing effect, while the linear
feedback dynamics \eqref{eq:5:2:1} is inherently unstable for
$K=101$. However, the stability analysis based on the
circle criterion given in Section \ref{sec:4}, does not deliver
conclusive results since one of the transformed closed
loops, cf. Figure \ref{fig:4:1:1}, is unstable. Nevertheless, an
additional stabilizing feature of hysteresis points to an
important effect in shaping the closed-loop system dynamics. This
does not appear to be controversial and can be interpreted as a local
stability property, while the absolute stability cannot be
guaranteed in this case. Local stability is supposedly related to the
amount of energy dissipated at the hysteresis cycles and is
therefore dependent on the hysteresis shape and initial
conditions in the otherwise unstable closed-loop dynamics. These
additional aspects of hysteresis stabilization are equally interesting,
but are beyond the scope of this study. They are certainly worth investigating
in the future.

\section{Conclusions}
\label{sec:6}

In this paper, we have addressed the stability of linear dynamics
with clockwise hysteresis in feedback. Starting by analyzing a
dissipative hysteresis map, we have demonstrated the property of
energy dissipation once the input changes and for each input
reversal. In this context, we have derived an upper boundary case
of a clockwise hysteresis, constructed with the help of a simple
sign operator applied to the input rate. In contrast, the lower
boundary case (where the hysteresis vanishes and the hysteresis
loops collapse to the static nonlinearity without memory), is
shown to be non-dissipative and therefore lossless from an energy
viewpoint. Other hysteresis shapes lying in-between the boundary
cases, i.e., having non-zero loop area, are shown to be equally
dissipative, cf. Figure \ref{fig:2:1}. A general class of linear
dynamic systems with hysteresis in feedback has been considered in
the spirit of the seminal work \cite{barabanov1979}. In addition
to existing approaches, a simple loop transformation which allows
for stability analysis by means of the circle criterion has been
introduced. It has been shown that the standard circle criterion
should be satisfied for both parallel loops of the transformed
dynamics, upon which the absolute stability can be concluded. The
numerical examples have been demonstrated for two boundary cases
of the second-order system dynamics. In addition, a more demanding
fourth-order dynamics of a feedback controlled double-mass
harmonic oscillator has been treated and discussed for one stable
and one unstable poles configuration. In this regard, some general
stabilizing hysteresis properties have also been discussed.

\bibliographystyle{plain}
\bibliography{references}

\end{document}